\documentclass[aps,pra,preprint,showpacs]{revtex4}
\usepackage{graphicx}
\usepackage{amssymb}
\usepackage{amsmath}

\newcommand{\htwo}{\mbox{H$_{2}$\,}}
\newcommand{\htwop}{\mbox{H$_{2}^{+}$\,}}

\DeclareMathAlphabet\mathbi{OML}{cmm}{b}{it}
\begin{document}
%------------------------------------------------
% TITLE, AUTHORS, AFFILIATION, ABSTRACT, AND PACS
\title{H$_2$ double ionization with few--cycle laser pulses}
\author{S. Saugout}
\affiliation{Laboratoire de Photophysique Mol\'{e}culaire du CNRS\\
Universit\'{e} Paris-Sud, b\^{a}timent 210, F--91405 Orsay, France}
\affiliation{Service des Photons, Atomes et Mol\'{e}cules\\
Direction des Sciences de la Mati\`{e}re,
CEA Saclay, b\^{a}timent 522, F--91191 Gif-sur-Yvette, France}
\author{E. Charron}
\affiliation{Laboratoire de Photophysique Mol\'{e}culaire du CNRS\\
Universit\'{e} Paris-Sud, b\^{a}timent 210, F--91405 Orsay, France}
\author{C. Cornaggia}
\affiliation{Service des Photons, Atomes et Mol\'{e}cules\\
Direction des Sciences de la Mati\`{e}re,
CEA Saclay, b\^{a}timent 522, F--91191 Gif-sur-Yvette, France}
%
%--------------
\begin{abstract}
The temporal dynamics of double ionization of H$_2$ has been
investigated both experimentally and theoretically with few--cycle
laser pulses. The main observables are the proton spectra associated
to the H$^+$~+~H$^+$ fragmentation channel. The model is based on
the time--dependent Schr\"odinger equation and treats on the
same level the electronic and nuclear coordinates. Therefore it
allows to follow the ultrafast nuclear dynamics as a function of the
laser pulse duration, carrier--envelope phase offset and peak
intensity. We mainly report results in the sequential
double ionization regime above $2 \times 10^{14}$~Wcm$^{-2}$.
The proton spectra are shifted to higher energies as the pulse duration
is reduced from 40~fs down to 10~fs. The good agreement between the
model predictions and the experimental data at 10~fs permits a theoretical
study with pulse durations down to a few femtoseconds. We demonstrate
the very fast nuclear dynamics of the H$_2^+$ ion for a pulse duration
as short as 1~fs between the two ionization events giving respectively
H$_2^+$ from H$_2$ and H$^+$~+~H$^+$ from H$_2^+$. Carrier--envelope
phase offset only plays a significant role for pulse durations shorter
than 4~fs. At 10~fs, the laser intensity dependence of the proton spectra
is fairly well reproduced by the model. 
\end{abstract}
\pacs{33.80.Rv, 33.80.Eh, 42.50.Vk}
\maketitle
%-------------
% INTRODUCTION
\section{INTRODUCTION}
Recent outstanding advances in ultrafast laser physics have led
to the generation of few--cycle pulses in the near--infrared and
visible ranges \cite{Nisoli_1996,Sartania_1997} and in the XUV range \cite{Paul_2001,Mairesse_2003}. Very efficient spectral techniques
are used for a complete reconstruction of the electric field in
the temporal domain. In the infrared and visible ranges, these
techniques are known as frequency--resolved optical gating (FROG)
\cite{Kane_1993} and spectral phase interferometry for direct
electric field reconstruction (SPIDER) \cite{Iaconis_1998}.
In the XUV range, the conventional optical elements used for
FROG and SPIDER cannot be used. In consequence,  the attosecond
pulse is reconstructed from spectral informations based on the
photoionization of atoms in the gas phase.
In the first attosecond experiments, the spectral phase was
measured through two--photon, two--color photoionization of
atoms \cite{Paul_2001,Mairesse_2003}. More recently, an
extension of the FROG concept to the XUV photoionization
was proposed and successfully applied to the measurement of
attosecond pulses \cite{Mairesse_2005,Sansone_2006}.

Following these developments, this report deals with the
temporal dynamics of double ionization of the H$_2$ molecule
induced by near--infrared few--cycle intense laser pulses in the
$10^{14}$--$10^{15}$~Wcm$^{-2}$ intensity range. The strong field
response of the H$_2$ molecule and its related molecular
ion and isotopic species have been extensively studied in the past
\cite{Giusti-Suzor_1995,Posthumus_2001,Posthumus_2004}. However, the
H$_2$--laser interaction remains a subject of great interest for
few--cycle laser pulses because of the attosecond electronic and
femtosecond nuclear time scales and the relatively
simple decay  channels (ionization and dissociation) leading to
H$_2^+$, H$^+$~+~H, and H$^+$~+~H$^+$. For instance in D$_2$,
the D$^+$~+~D channel was used to propose the idea of
an attosecond molecular clock based on the rescattering
dynamics of the first ionized electron leading to the
dissociation of the D$_2^+$ ion \cite{Niikura_2003}.
The same excitation scheme and dissociation channel were
recently studied in order to control the electron localization
with 5--fs carrier--envelope--phase--locked laser pulses
\cite{Kling_2006}. We have recently shown that the proton spectrum
allows to detect the presence of a pre--pulse or a post--pulse
using a 10--fs pump--probe scheme \cite{Saugout_2006} and the
important concept of charge resonance enhanced ionization
introduced by Bandrauk \textit{et al.} \cite{Zuo_1995}.

In H$_2$, double ionization leads to two bare protons.
In consequence, our main experimental and theoretical diagnostic
is the proton spectrum as a function of the pulse duration,
carrier--envelope offset phase, and intensity. 
Double ionization induced by few--cycle laser pulses exhibit
two regimes : nonsequential double ionization at low laser
intensities below $2 \times 10^{14}$~Wcm$^{-2}$ and sequential
ionization at higher laser intensities
\cite{Legare_2003,Tong_2003,Tong_2004,Alnaser_2004,Rudenko_2005,
Beylerian_2006}. Charge resonance enhanced ionization belongs to
the sequential ionization regime. However, it does not occur
for few--cycle laser pulses because of the ultrashort pulse
duration. After the first ionization event of H$_2$, the resulting 
H$_2^+$ molecular ion does not have enough time to stretch and
to reach the internuclear distance range where enhanced ionization
takes place \cite{Zuo_1995}. Nonsequential double ionization has
been studied in detail with an emphasis on the rescattering dynamics
\cite{Tong_2003,Alnaser_2004}. Here we propose an experimental
and theoretical study of the sequential regime at intensities
above $2 \times 10^{14}$~Wcm$^{-2}$ where rescattering is less
important \cite{Tong_2004}. The interesting feature is
the time delay between the first and second electron removals.
For few--cycle laser pulses this time delay might be expected to
be a few femtoseconds corresponding to the pulse risetime.
In the meantime, a nuclear wave packet arises in the
H$_2^+$ molecular ion from the nonresonant coupling of the
ground electronic state X$^2\Sigma_g^+$ and the first excited
dissociative state A$^2\Sigma_u^+$. Therefore the delayed
ejection of the second electron takes place during the evolution
of the molecular ion internuclear distance. The resulting proton
spectrum is shifted to lower energies in comparison to what might
be expected from an instantaneous two--electron ejection.

In spite of the simplicity of the above picture , a quantitative
prediction of the proton spectrum as a function of the ultrashort
pulse parameters is not straightforward because of the complicated
nonlinear couplings leading to single and double ionizations.
In addition, the evolution of the nuclear wave packet of the
intermediate H$_2^+$ ion has to be included in the theoretical
framework. Usually this problem is solved using a two--step approach
\cite{Legare_2003,Tong_2004}. After the first ionization event,
the nuclear evolution of the H$_2^+$ ion is solved by numerical
integration of the time--dependent Schr{\"o}dinger equation.
The initial wave packet is given by the projection of the
H$_2$ ground vibrational state onto the different ion vibrational
states. The second ionization event leading to double ionization
is then calculated as a function of the time--dependent vibrational
wave packet. Tong and Lin give a clear account of the procedure
in Ref.~\cite{Tong_2004}. Here we propose a complementary approach for
the quantitative analysis of our experimental results. Double
ionization is treated using a unified approach based on a 
two--electron model where the internuclear distance remains a
full quantum variable in order to extract the nuclear dynamics
during the interaction. This model has been used in a recent paper
\cite{Saugout_2007} for the study of the mechanisms leading to double
ionization with 1--fs laser pulses. This manuscript focuses on other
effects and on a more detailed comparison with experiments.

The paper is organized as follows. The experimental set--up is
presented in Section~\ref{section_exptal} including the 10--fs pulses
generation set--up and the time--of--flight detection of the ions.
The theoretical model is described in Section~\ref{sec:Theory} with
an emphasis on a realistic field--free molecular description.
In Section~\ref{section_results}, the experimental results
are compared to the theoretical predictions for a better understanding
of the ionization and fragmentation mechanisms.
In Section~\ref{section_results}, pulse duration, carrier--envelope
phase, and intensity dependences of the proton spectra are successively
presented and discussed. Since the carrier--envelope phase of our
10--fs pulses is not controlled experimentally, this dependence will
be commented from theoretical predictions only. The conclusions are
finally summarized in the last section.

%--------------------
% EXPERIMENTAL SET--UP
%
\section{EXPERIMENTAL SET--UP}
\label{section_exptal}
\subsection{Laser system and pulse compression method}
The ultrashort pulse generation set--up is based on a 1--kHz
titanium:sapphire laser system and a hollow fibre pulse
compression stage. The 1--kHz laser chain is built following
the conventional chirp--pulse--amplification scheme
\cite{Strickland_1985}. It consists of an oscillator, a stretcher,
a regenerative amplifier, and a compressor. The system delivers
pulses with energies up to 600~$\mu$J, duration of 40~fs, and
a central wavelength of 795 nm. The pulse compression stage is
designed following techniques introduced by
Nisoli \textit{et al.} \cite{Nisoli_1996} and
Sartania \textit{et al.} \cite{Sartania_1997}.
The laser beam is focused with a 700--mm--focal--length lens
onto the tip of a 700--mm--long hollow fibre with a
250--$\mu$m inner diameter. The hollow fibre is housed
on a V--groove in a chamber filled with argon gas.
The non--linear Kerr effect in argon leads to self--phase
modulation and wavelength spectrum broadening while the fibre
waveguide ensures a spatially--homogeneous spectral broadening.
Optimum argon operating pressures were found around 700~mbar
for 40--fs and 600--$\mu$J input laser pulses.    
%
%------------------------
% Figure 1 : laser spectra
\begin{figure}[htbp]
\begin{center}
\includegraphics*[width=0.5\textwidth]{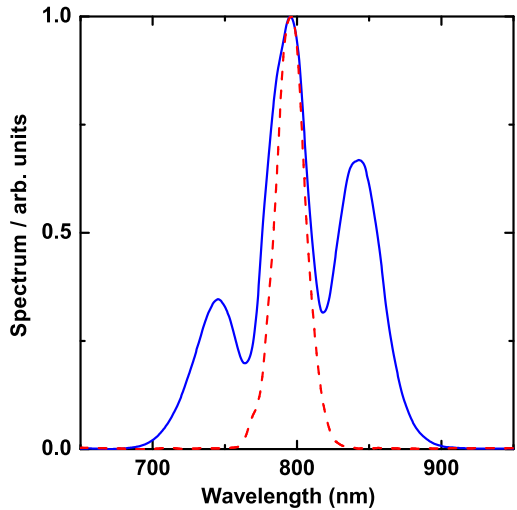}
\end{center}
\caption{(Color online) Wavelength spectra before (dashed curve)
and after the hollow fibre (full curve) filled with argon gas.
The argon pressure is 700~mbar. The laser pulse energy and
duration are respectively 600~$\mu$J and 40~fs.}
\label{fig_laser_spectra}
\end{figure}
%--------------------------

Figure~\ref{fig_laser_spectra} shows the wavelength spectra
before and after the hollow fibre filled with argon. Self--phase
modulation leads to a noticeable broadening. The usual multi--peak
structure is observed as in other studies
\cite{Nisoli_1996,Sartania_1997}. The oscillatory behavior comes
from interferences of waves with the same frequency but generated
at different times within the laser pulse. The overall transmission
of the hollow fibre set--up including all the optical elements is
above 50~$\%$. After recollimation by an $f=1$~m concave silver
mirror, pulses are recompressed to 10~fs using several reflections
on commercial broadband chirped mirrors. The pulse duration is
measured using a home--made interferometric autocorrelator.
%
%--------------------------
% Figure 2 : autocorrelation
\begin{figure}[htbp]
\begin{center}
\includegraphics*[width=0.5\textwidth]{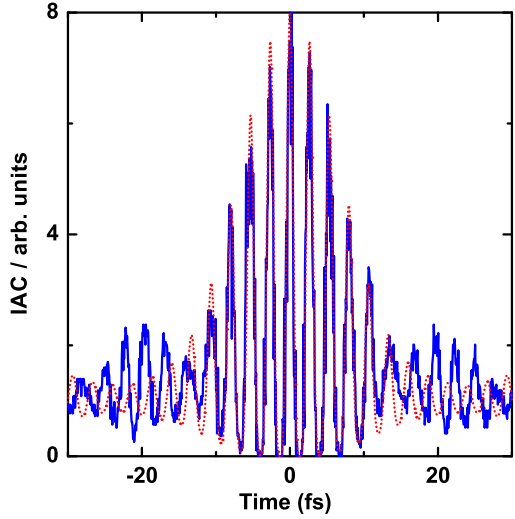}
\end{center}
\caption{(Color online) Interferometric autocorrelation (IAC) of
10--fs laser pulses. The experimental and calculated signals are
given respectively by the full and dotted curves. The IAC curve is
calculated using the Fourier transform of the frequency domain
electric field, assuming a constant spectral phase.}
\label{fig_laser_autocos}
\end{figure}
%-----------
%
Figure~\ref{fig_laser_autocos} presents an interferometric
autocorrelation of 10--fs pulses corresponding to the wavelength
spectrum presented in Fig.~\ref{fig_laser_spectra}. The measured
autocorrelation signal is compared to a calculated
interferometric autocorrelation assuming a constant spectral
phase. The good agreement between the measured and calculated
curves shows that the chirped--mirrors compression stage works
well for the second--order group delay dispersion compensation.
The remaining disagreement comes from cubic and quartic residual
phase that cannot be compensated with our set--up. The laser pulse
duration relative uncertainty is estimated to be $\pm20$~\%.
Ultrashort 10--fs pulses with energies above 200~$\mu$J are
therefore available for subsequent experiments.

The pulse duration measurement presented in
Fig.~\ref{fig_laser_autocos} has been recorded after the
optimization of the group delay dispersion. This optimization is
only valid for the autocorrelator because the chirped mirrors
compressor is tuned in order to compensate for the air travel to the
autocorrelator and also for the small amount of group delay dispersion
introduced by this device. The proton spectra are recorded in 
a vacuum chamber which is located elsewhere in the laboratory
and which introduces a different group delay dispersion than the
autocorrelator. Therefore the pulse duration will have to be optimized
\textit{in situ} at the location where the laser--molecule
interaction takes place. The method is to introduce more negative
group delay dispersion than necessary and then to compensate for it
with a variable thickness of fused silica which exhibits a positive
dispersion of 36.1~fs$^2$rad$^{-1}$mm$^{-1}$. Proton spectra are
systematically recorded for different fused silica thicknesses.
As expected, the overall proton spectrum is shifted to the
highest energy for the shortest pulse duration.
This optimization procedure will be commented in more detail
in Section~\ref{section_results} since it involves a thorough
understanding of the molecular response.

%---------------------------------
\subsection{Ion detection set--up}
%
%---------------------
% Figure 3 : covariance
\begin{figure}[htbp]
\begin{center}
\includegraphics*[width=0.5\textwidth]{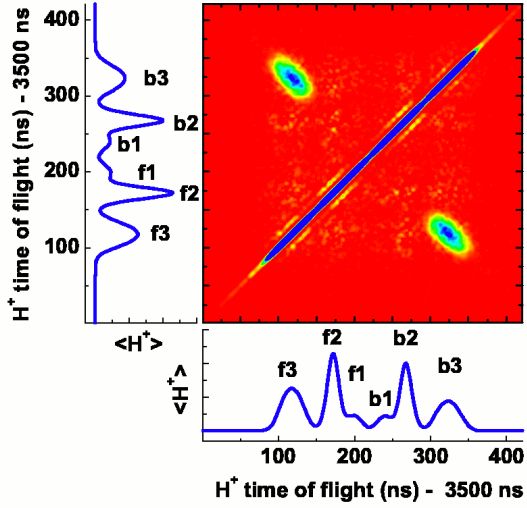}
\end{center}
\caption{(Color online) Covariance map from H$_2$ recorded with
linearly--polarized 40--fs laser pulse at
$I = 2 \times 10^{14}$~Wcm$^{-2}$ and
$p(\textrm{H}_2)= 5 \times 10^{-8}$~mbar. The collection electric
field is 25~Vcm$^{-1}$. The bottom and left panels represent the
usual proton time--of--flight spectrum and the central panel
represents the covariance map. The f1, f2, and f3 peak
labels correspond to protons ejected towards the detector.
The b1, b2, and b3 labels are associated to protons ejected
backwards the detector.}
\label{fig_covariance}
\end{figure}
%----------
The ultrashort pulses are sent in an ultrahigh vacuum chamber
equipped with a 75--mm--focal--length on--axis parabolic mirror
which allows to get laser intensities up to $10^{16}$~Wcm$^{-2}$. 
The hydrogen gas is introduced through an effusive gas jet at
very low pressures down to $3 \times 10^{-10}$~mbar which is
the residual pressure of the chamber. Molecular hydrogen ions
and protons are detected using a
1150--mm--long time--of--flight spectrometer based on the
Wiley--McLaren configuration and devoted to experimental
studies of multiple ionization \cite{Wiley_1955,Baldit_2005}.
Fragmentation channels and the associated kinetic energy release
spectra are determined using covariance mapping introduced by
Frasinski \textit{et al.} \cite{Frasinski_1989,Hering_1999}.
In particular for H$_2$, it is important to separate the
H$^+$~+~H$^+$ double ionization channel from the H$^+$~+~H
single ionization channel. 

We recall here the main features of this
technique. First of all, it allows to work with more than one
ionization event per laser shot. Fragments coming from the same
dissociation channel are expected to produce time--of--flight
signals that fluctuate following a same pattern on a
shot--to--shot basis. For fragmentation channels involving two
detected bodies, the covariance coefficient $C_2(T_1,T_2)$
therefore measures the statistical correlation of
fragments arriving at times $T_1$ and $T_2$. The method is
illustrated in Fig.~\ref{fig_covariance}. The conventional
time--of--flight proton spectrum is represented in the bottom
panel. The time--of--flight signal has been delayed by 3500~ns in
order to record only the significant proton signals. The proton
spectrum is highly symmetric around the relative time of flight
220~ns and exhibits 3 peaks of protons emitted towards the
detector and labeled f1, f2, and f3. The peaks labeled
b1, b2, and b3 are associated to protons emitted backwards
the detector. Let us recall here that the proton time of flight $T$
is given by $T = T_0 \pm P/eF_c$, where $T_0$ is the time of
flight of a proton with a zero initial momentum, $P$ is the
modulus of the initial momentum, the $\pm$ sign is positive for
a proton emitted away from the detector and negative for a proton
emitted towards the detector, $F_c$ is the collection electric field
($F_c = 25$~Vcm$^{-1}$ in these experiments) and $e$ is the elementary
charge \cite{Hering_1999}. In addition our spectrometer exhibits a
strong angular discrimination due to its large longitudinal dimension.
Therefore it allows the detection of protons with an initial momentum
parallel to its axis. Since the angular discrimination is stronger for
backwards protons, the backwards peaks b1, b2, and b3 in
Fig.~\ref{fig_covariance} are slightly smaller than the f1, f2, and f3
peaks associated to forwards protons.

%---------------------------
% Figure 4 : protons D1 et D2
\begin{figure}[htbp]
\begin{center}
\includegraphics*[width=0.5\textwidth]{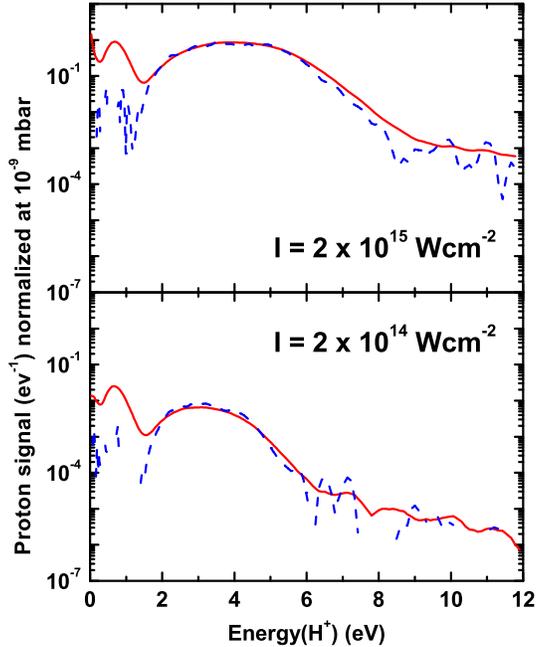}
\end{center}
\caption{(Color online) Proton spectra from H$_2$ recorded with
linearly--polarized 40--fs laser pulse at
$I = 2 \times 10^{14}$~Wcm$^{-2}$ and 
$I = 2 \times 10^{15}$~Wcm$^{-2}$.
For each laser intensity, the full curve represents the total proton
spectrum and the dashed curve represents the covariance proton
spectrum from the H$^+$~+~H$^+$ dissociation channel.}
\label{fig_proton_D1D2}
\end{figure}
%-----------
%
The covariance coefficient $C_2(T_1,T_2)$ is represented in the
central map in Fig.~\ref{fig_covariance}.
For an easy visualization of the correlations, the
conventional proton time--of--flight spectrum is also represented
in the vertical left panel. A correlation between peaks f3 and
b3 clearly appears in Fig.~\ref{fig_covariance} and corresponds to
the Coulomb explosion channel H$^+$~+~H$^+$. The peaks f1 and f2
and their associated backwards peaks b1 and b2 are not correlated
with any other peaks. They are associated to the H$^+$~+~H
dissociation channel. Following the relationship between the
time of flight and the momentum, the proton kinetic energy
release spectra are extracted from the time--of--flight data
using the associated time--to--energy normalization factors.
Some proton energy spectra are represented in
Fig.~\ref{fig_proton_D1D2}. The bottom curves represent the total
and the H$^+$~+~H$^+$ covariance energy spectra extracted from
the time--of--flight data presented in Fig.~\ref{fig_covariance}.
Although there remains some covariance noise for protons coming
from H$^+$~+~H, the proton spectrum associated to the H$^+$~+~H$^+$
channel is unambiguously identified as the broad peak above 1.5~eV
of the conventional proton energy spectrum. Therefore in the
following, we only present conventional proton spectra which
exhibit a better statistics than the covariance spectra. 

%-------
% THEORY
%
\section{Theoretical Treatment}
\label{sec:Theory}
This section introduces the theoretical background and numerical
methods we have developed for studying strong field Coulomb
explosion dynamics of the hydrogen molecule.

% Molecular system
\subsection{The molecular system and the laser--molecule interaction}
\label{subsec:Theo-model}
We follow the double ionization dynamics of the \htwo molecule by
solving the time--dependent Schr\"odinger equation for the electronic
and nuclear motions
\begin{equation}
\label{eq:TDSE}
i\hbar\,\frac{\partial}{\partial t} \chi_{s}\Psi(\mathbi{R},\mathbi{r},t) =
\left[\hat{\cal{H}}_0 + V_{\mathrm{int}}\right] \chi_{s}\Psi(\mathbi{R},\mathbi{r},t)\,,
\end{equation}
where $\hat{\cal{H}}_0$ is the field--free Hamiltonian and $V_{\mathrm{int}}$
the field--molecule interaction potential. The body-fixed coordinate
$\mathbi{r} \equiv \left\{\mathbi{r}_1,\mathbi{r}_2\right\}$ refers here
to the electrons, while $\mathbi{R}$ represents the internuclear vector.
The electronuclear wave packet is denoted by
$\chi_{s}\Psi(\mathbi{R}, \mathbi{r}, t)$, where $\chi_{s}$ is an antisymmetric
two--electron spin wave function, while the spatial wave function
$\Psi(\mathbi{R}, \mathbi{r}, t)$ is symmetric with respect to the exchange
of the two electrons (singlet state). This approach does not assume a
separation of the electronic and nuclear coordinates, thus going beyond
the usual Born-Oppenheimer approximation.

The Hamiltonian of the field--free diatomic molecule is expressed as
\begin{equation}
\label{eq:H0}
\hat{\cal{H}}_0 = \hat{T}_{n} + \frac{1}{R} +
                  V_{12}\left(\mathbi{r}_1-\mathbi{r}_2\right) + \sum_{i=1,2}\hat{h}_{i}\,,
\end{equation}
where $\hat{T}_{n}$ is the nuclear kinetic operator and
$V_{12}(\mathbi{r}_1-\mathbi{r}_2)$ the inter--electronic repulsion. The mono--electronic Hamiltonians $\hat{h}_{i}$ are expressed as the sum of the electronic kinetic operator and the electron--nuclei interaction potential
\begin{equation}
\label{eq:he}
\hat{h}_{i} = -\frac{\hbar^2}{2m}\nabla^{2}\!\!\!_{i} + V_{en}(\mathbi{R}, \mathbi{r}_i)\,.
\end{equation}

The ionization dynamics is initiated by the length gauge radiative coupling
\begin{equation}
\label{eq:Vint}
V_{\mathrm{int}}(\mathbi{r}_1,\mathbi{r}_2,t)=-e\left(\mathbi{r}_1+\mathbi{r}_2\right)\cdot\mathbi{E}(t)
\end{equation}
associated, in the dipole approximation, with the linearly polarized classical electric field
\begin{equation}
\label{eq:field}
\mathbi{E}(t)= E_{0}\,f(t)\cos(\omega t + \varphi)\;\hat{\mathbi{\!e}}\,,
\end{equation}
where $E_{0}$ is the field amplitude, $\omega$ the angular frequency,
and $\varphi$ the carrier--envelope offset phase. The pulse shape is
given by the Gaussian--like expression
\begin{equation}
\label{eq:pulse}
f(t)=\sin^{2}\left(\frac{\pi t}{\tau}\right)\,,
\end{equation}
with total pulse duration $\tau$. The frequency $\omega$ corresponds to
a central wavelength of $800$\,nm and the internuclear coordinate
$\mathbi{R}$ is constrained along the field polarization vector $\hat{\mathbi{\!e}}$. The two electrons, of coordinates $\mathbi{r}_i=z_i\;\hat{\mathbi{\!e}}$, are also assumed to
oscillate along this axis. Recent numerical studies have indeed
shown that two-electron dynamics in molecules is mainly characterized
by a one-dimensional motion in linear polarization\,\cite{Harumiya_2002,Becker_2006}.

\begin{figure}[htbp]
\begin{center}
\includegraphics*[width=0.5\textwidth]
                 {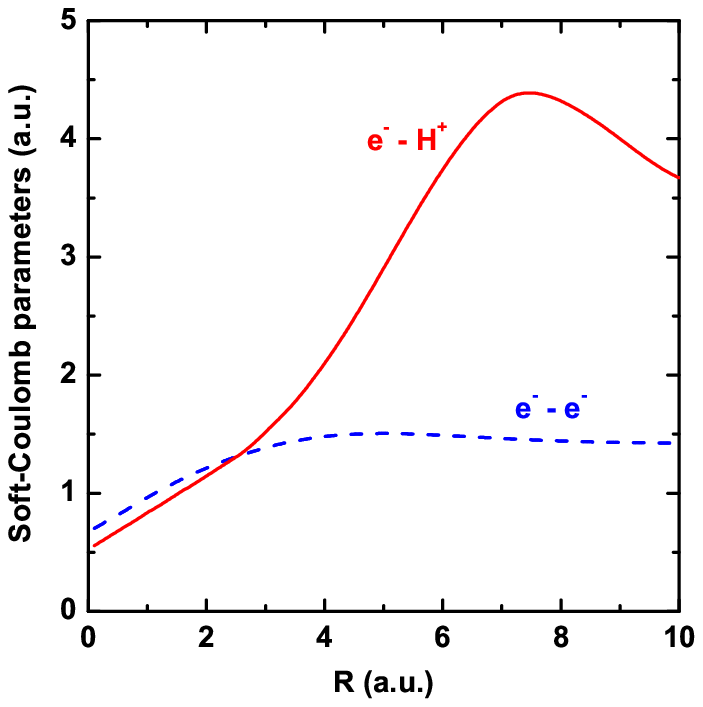}
\end{center}
\caption{(Color online) Soft--Coulomb parameters. The full curve
and the dashed curves represent respectively the
electron--nuclei and the electron--electron
smoothing parameters.}
\label{fig_H2_initial_parameters}
\end{figure}

In order to mimic the dynamics of the real \htwo molecule, we have
introduced two $R$--dependent softening parameters
$\alpha(R)$ and $\beta(R)$ in the Coulomb potentials describing the electron--electron
\begin{equation}
\label{eq:Vee}
V_{12}(z_1-z_2) = \left[(z_1 - z_2)^2 + \alpha^2(R)\right]^{-\frac{1}{2}}
\end{equation}
and electron-nuclei interactions
\begin{equation}
\label{eq:Ven}
V_{en}(R, z_{i}) = - \sum_{s = \pm 1} \left[(z_{i} + s \times R/2)^2 + \beta^2(R)\right]^{-\frac{1}{2}}\,.
\end{equation}
%
%---------------------------------------------------------
% Figure 5 : potential curves of initial H$_2$ and H$_2^+$
\begin{figure}[htbp]
\begin{center}
\includegraphics*[width=0.5\textwidth]
                 {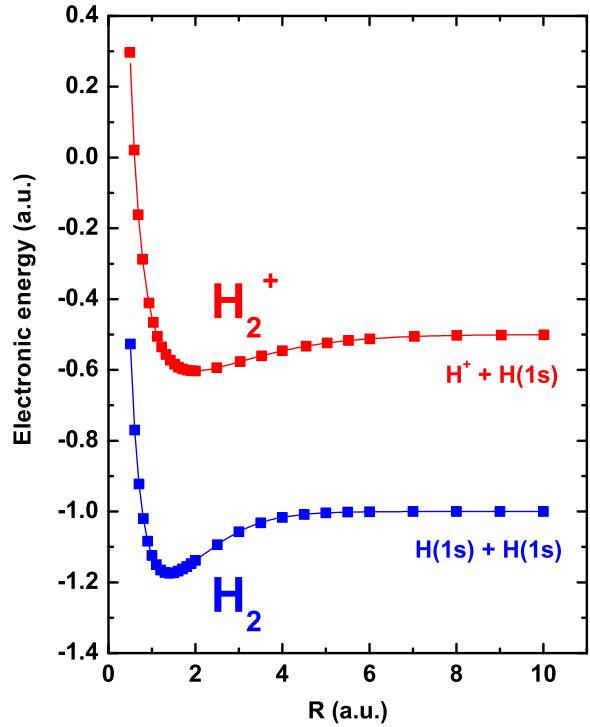}
\end{center}
\caption{(Color online) Calculated potential curves of the
electronic ground states of H$_2$ and H$_2^+$. Squares are
obtained from the model presented in the paper and full curves
are the \textit{ab initio} energies reported in
Refs.~\cite{Kolos_1965, Peek_1965}.}
\label{fig_H2_initial_energies}
\end{figure}
%-----------
%
These two softening parameters are assumed to vary slowly with
the internuclear distance. In a first step, the parameter
$\beta(R)$ is adjusted at each internuclear distance in order
to reproduce the energy of the ground electronic state of
\htwop\,\cite{Peek_1965}. The parameter $\alpha(R)$ is then
obtained by reproducing the energy of the ground electronic
state of the hydrogen molecule\,\cite{Kolos_1965}. These
potential curves, presented in Fig.~\ref{fig_H2_initial_energies},
confirm that the adjustment of these two parameters allows for an
accurate reproduction of the exact molecular potentials even though
the electronic problem is presently reduced to a single dimension.
The variation of these two parameters with the internuclear
distance is given in Fig.~\ref{fig_H2_initial_parameters}.
%
%-----------------------------------
% Figure 6 : soft Coulomb parameters
%-----------

% Electronuclear wave packet
\subsection{The electronuclear wave packet propagation and the
initial \htwo wavefunction}
\label{subsec:Theo-propag}
In order to calculate the single and double ionization of the hydrogen
molecule submitted to an intense and pulsed laser radiation,
we propagate the total wave function $\Psi(\mathbi{R},\mathbi{r},t)$
in time during the entire pulse using the split operator method
developed by Feit et al.\,\cite{Feit_1982}
\begin{equation}
\label{eq:propag}
\Psi(\mathbi{R},\mathbi{r},t+\delta t) = \mathrm{e}^{-i\,\left[\hat{\cal{H}}_0 + V_{\mathrm{int}}\right]\,\delta t/\hbar}\;\Psi(\mathbi{R},\mathbi{r},t)\,,
\end{equation}
where the total Hamiltonian $\hat{\cal{H}}_0 + V_{\mathrm{int}} = \hat{T}+\hat{V}(t)$ is split in two parts corresponding to the
kinetic and potential propagations
\begin{eqnarray}
\label{eq:split}
\mathrm{e}^{-i\,\left[\hat{\cal{H}}_0 + V_{\mathrm{int}}\right]\,\delta t/\hbar}
& = &
\mathrm{e}^{-i\,\hat{T}\,\delta t/2\hbar}
\;
\mathrm{e}^{-i\,\hat{V}(t)\,\delta t/\hbar}\nonumber\\
& &
\times \mathrm{e}^{-i\,\hat{T}\,\delta t/2\hbar}
+o(\delta t^3)\,.
\end{eqnarray}
$\hat{T}$ represents here the sum of the nuclear and electronic
kinetic energy operators, while $\hat{V}(t)$ includes all potential
operators. The kinetic and potential propagations are performed
in the momentum and coordinate spaces respectively. Three-dimensional
Fast Fourier Transformation (FFT) allows rapid passage back and forth
from one representation to the other at each time step. Typical grids
extend up to $z_1^{\mathrm{max}}=z_2^{\mathrm{max}}=100\,a_0$ and $R^{\mathrm{max}}=10\,a_0$ with $(512)^3$ grid points. A time step of
$\delta t \simeq 1\,$as is necessary to achieve convergence. For very
short pulse durations, an additional field free propagation is performed
after the end of the pulse in order to give enough time for the ionized
electrons to reach the asymptotic region were the wave function can be
analyzed (see Sec.\,\ref{subsec:Theo-analyse} hereafter).

%-----------------------------------------
% Figure 7 : vibration of initial H_2, v=0
\begin{figure}[htbp]
\begin{center}
\includegraphics*[width=0.5\textwidth]
                 {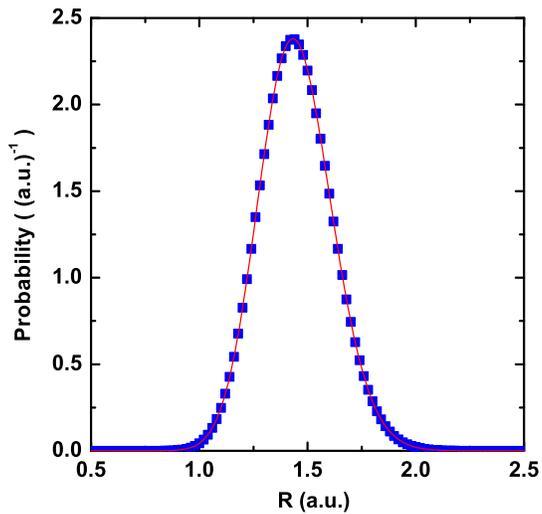}
\end{center}
\caption{(Color online) Probability density from the vibrational
ground state of H$_2$.
Full curve: This model. 
Squares: Calculated from the \textit{ab initio} ground electronic
state $v=0$ of H$_2$ given in Ref.~\cite{Kolos_1965}.}
\label{fig_H2_initial_vibration}
\end{figure}
%-----------
The initial wave function $\Psi(\mathbi{R},\mathbi{r},t=0)$ is taken
as the ground nuclear and electronic state of the hydrogen molecule,
calculated within the present non--Born--Oppenheimer model using the
imaginary time relaxation technique\,\cite{Kosloff_1986}. The equilibrium
internuclear distance of \htwo, $R_{\mathrm{e}} \simeq 1.4\,a_0$, is perfectly
reproduced, and Fig.~\ref{fig_H2_initial_vibration} shows the very good agreement
obtained between the nuclear probability density
\begin{equation}
\label{eq:PR}
P(R) = \iint \left| \Psi(\mathbi{R},\mathbi{r},t=0) \right|^2\, dz_{1} \, dz_{2}\,,
\end{equation}
calculated here and the probability density $\chi_v^2(R)$ of the ground vibrational
state $v=0$ of \htwo calculated from the Born--Oppenheimer ground electronic state
potential given in Ref.~\cite{Kolos_1965}.

% Wave packet analysis
\subsection{The wave packet analysis}
\label{subsec:Theo-analyse}
\begin{figure}[htbp]
\begin{center}
\includegraphics*[width=0.5\textwidth]
                 {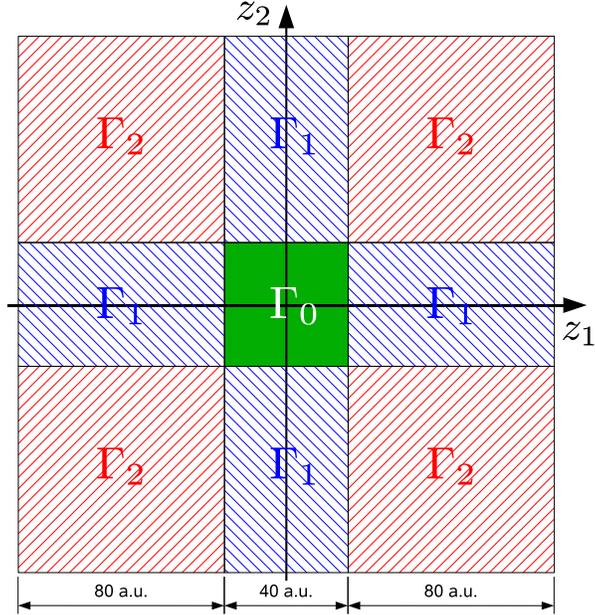}
\end{center}
\caption{(Color online) Partition zones of the electrons coordinates
$z_1$ and $z_2$. The three regions $\Gamma_0$, $\Gamma_1$ and $\Gamma_2$
correspond respectively to the
H$_{2}$,
H$_{2}^{+} + e^-$, 
and H$_{2}^{2+} + e^- + e^-$ 
systems.}
\label{fig_H2_initial_zones}
\end{figure}

The single and double ionization probabilities are analyzed using a
well--established cartography technique\,\cite{Pegarkov_1999}.
The plane $(z_1,z_2)$ is partitioned in three regions
$\Gamma_0$, $\Gamma_1$ and $\Gamma_2$ corresponding respectively
to H$_{2}$, H$_{2}^{+}$, and H$_{2}^{2+}$. Double ionization occurs
in the asymptotic region $\Gamma_2 \equiv \{|z_1|,|z_2|>z_{\mathrm{A}}\}$
with $z_{\mathrm{A}} = 20\,a_0 \gg R_e$. The single ionization region
$\Gamma_1$ is defined as $\{|z_i|<z_{\mathrm{A}}$, $|z_j|>z_{\mathrm{A}}\}$,
and the neutral H$_{2}$ molecule is found in the region
$\Gamma_0\equiv\{|z_1|,|z_2| \leqslant z_{\mathrm{A}}\}$. A schematic
illustration of these three regions is given in
Fig.~\ref{fig_H2_initial_zones} which represents the partition of the
$(z_1, z_2)$ plane. The outgoing ionization flux is accumulated during
the time propagation in the regions $\Gamma_1$ and $\Gamma_2$ to extract
the single and double ionization probabilities as a function of time.
Absorbing boundaries are imposed at the end of the grid to avoid spurious
reflection effects.
%
%---------------------------
% Figure 8 : partition zones
%----------

To extract the proton kinetic energy distributions $S(E,t)$ obtained
after Coulomb explosion, we use a simple mapping which relates $S(E,t)$
to the probability density
\begin{equation}
\label{eq:PR2}
P_2(R,t) = \iint_{\Gamma_2} \left| \Psi(\mathbi{R},\mathbi{r},t) \right|^2 \, dz_{1} \, dz_{2}\,,
\end{equation}
in the $\Gamma_2$ region, using the Coulomb relation $E=0.5/R$ and
the requirement of particle conservation
\begin{equation}
\label{eq:PC}
P_2(R,t)\,dR=S(E,t)\,dE\,.
\end{equation}
The kinetic energy $E$ denotes here the energy of a single proton.
The energy distribution $S(E,t)$ is finally accumulated over the
entire time propagation to obtain the kinetic energy release spectrum
$S(E)$ which is measured in experiments.

%-----------------------------------------
% Figure 9 : comparison at 10 fs and 12 fs
\begin{figure}[b]
\begin{center}
\includegraphics*[width=0.5\textwidth]
                 {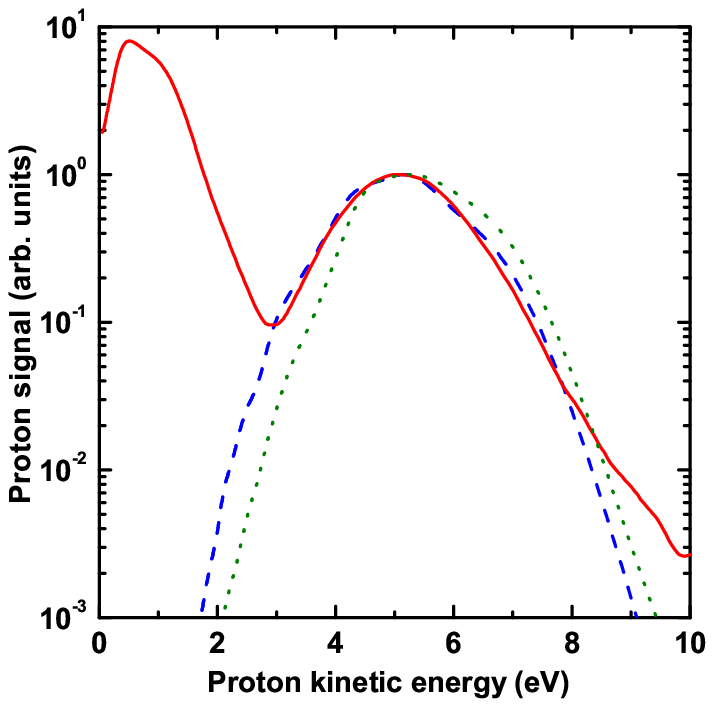}
\end{center}
\caption{(Color online) Proton spectra from H$_2$ at
$4 \times 10^{14}$~Wcm$^{-2}$. 
Full curve: Experimental with a pulse duration of 10~fs.
Dotted curve: Theoretical with a pulse duration of 10~fs.
Dashed curve: Theoretical with a pulse duration of 12~fs.}
\label{fig_proton_exp_th_12fs_10fs}
\end{figure}
%----------
%
Figure~\ref{fig_proton_exp_th_12fs_10fs} represents the comparison
between an experimental proton spectrum recorded at
$4 \times 10^{14}$~Wcm$^{-2}$ with a pulse duration of 10~fs and
two spectra calculated at the same laser intensity with
pulse durations of respectively 10~fs and 12~fs. The proton
peak associated to the H$^+$~+~H$^+$ channel is better reproduced
with a 12--fs pulse duration than with a 10--fs duration. It is
quite noticeable that a relatively small variation of 2~fs leads
to a measurable proton energy shift. Concerning the agreement
with the experimental data, the very high sensitivity of the
proton spectra might explain the discrepancy between the
experimental and theoretical proton spectra at 10~fs when one
takes into account the relative accuracy of the pulse duration
and laser intensity measurements. In the following, we choose
to compare the experimental data with calculations performed
with a pulse duration of 10~fs in order to deal with a
theoretical model without any adjustable parameter. The
agreement at different laser intensities will be commented
more thoroughly in Section~\ref{section_results}.

%-----------------------
% RESULTS AND DISCUSSION
%
\section{RESULTS AND DISCUSSION}
\label{section_results}
\subsection{Pulse duration dependence of proton spectra}
%
%---------------------------------------------
% Figure 10 : proton spectra at 40 fs and 10 fs
\begin{figure}[b]
\begin{center}
\includegraphics*[width=0.5\textwidth]{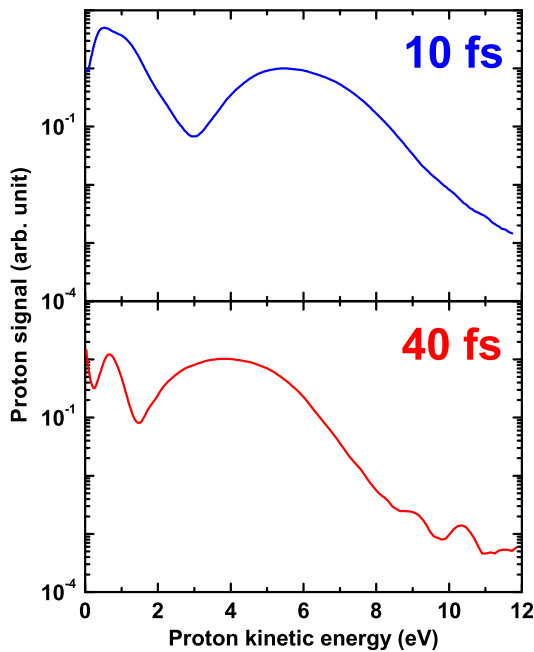}
\end{center}
\caption{(Color online) Proton spectra from H$_2$ at
$8 \times 10^{14}$~Wcm$^{-2}$ recorded with pulse durations
of 40~fs and 10~fs.}
\label{fig_proton_40fs10fs}
\end{figure}
\begin{figure}[htbp]
\begin{center}
\includegraphics*[width=0.5\textwidth]{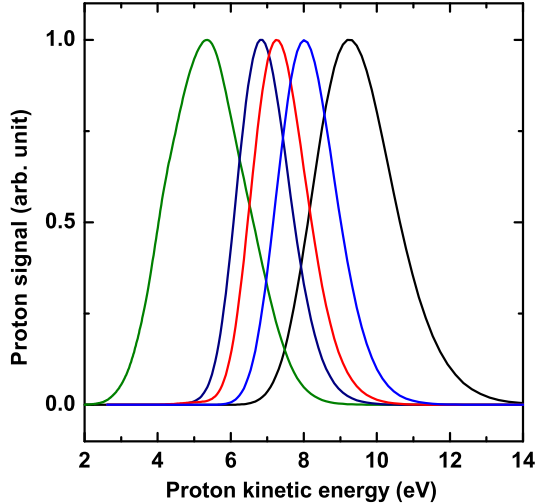}
\end{center}
\caption{(Color online) Proton spectra from H$_2$ calculated at
$8 \times 10^{14}$~Wcm$^{-2}$ for different pulse durations.
From right to left: ``0'', 1, 2, 4, and 10~fs. The ``0''--fs
pulse duration spectrum represents what is expected from an
instantaneous two--electron ionization of H$_2$.}
\label{fig_proton_0a11CO}
\end{figure}

%-----------
%
This section is devoted to the dependence of the proton spectra
as a function of the pulse duration. The comparisons between
different pulse durations are performed while keeping the same
peak laser intensity.
For instance, Fig.~\ref{fig_proton_40fs10fs} represents two
proton spectra recorded at $8 \times 10^{14}$~Wcm$^{-2}$ with
respectively 40--fs and 10--fs laser pulses. Both spectra
exhibit the above--mentioned separation between the
H$^+$~+~H and H$^+$~+~H$^+$ dissociation channels. In particular
for this last channel, the proton spectra have maxima at
3.8~eV at 40~fs  and 5.5~eV at 10~fs. We first would like to
emphasize the large shift of 1.7~eV when the pulse duration
is reduced from 40~fs to 10~fs.
It is also important to notice that double ionization
cannot be considered as an instantaneous double ionization
process at 10~fs since such a sudden electrons removal would
produce a proton spectrum peaked at 9.2~eV. This instantaneous
two--electron emission would lead to the proton spectrum
represented in Fig.~\ref{fig_proton_0a11CO} by the far
right curve. The corresponding calculation is based on the
projection of the population of the $v=0$ vibrational state 
of H$_2$ represented in Fig.~\ref{fig_H2_initial_vibration}
onto the H$^+$~+~H$^+$ repulsion curve with the appropriate
normalization factors.
The interpretation is here straightforward: After the first
ionization step, enough time is left to the H$_2^+$ ion for
a significant stretching before the second ionization event.
The proton peak shifts towards higher energies at 10~fs since
the internuclear distance is reduced because of the shorter pulse
duration.

%----------------------------------------------------
% Figure 11 : proton spectra at 0, 1, 2, 4, and 10 fs
%---------- 
To the best of our knowledge, intense infrared laser pulses with
durations below a few femtoseconds are not yet available.
Theoretical predictions thus become highly desirable in order
to know the H$_2$ molecule sensitivity to ultrashort pulses.
Figure~\ref{fig_proton_0a11CO} shows five calculated proton
spectra from right to left which correspond to the instantaneous
two--electron ionization for the far right spectrum and then
to pulse durations of respectively 1~fs, 2~fs, 4~fs, and 10~fs.
The laser peak intensity remains $8 \times 10^{14}$~Wcm$^{-2}$
as in the experimental results in Fig.~\ref{fig_proton_40fs10fs}.
All curves have been normalized to unity for an easier comparison.
The calculations are performed with a zero carrier--envelope
offset phase. The most striking feature comes from the proton
spectrum calculated with a 1--fs pulse duration. The spectrum
peaks at 8~eV and is therefore already shifted by 1.2~eV towards lower
energies in comparison with the proton spectrum from instantaneous
double ionization. Even for such an ultrashort laser pulse,
nuclear motion takes place and leads to a measurable shift
of the proton spectrum. As for the experimental spectra
presented in Fig.~\ref{fig_proton_40fs10fs}, the theoretical
proton spectra are shifted to lower energies as the pulse
duration is increased from 1~fs to 10~fs. The spectrum calculated et
10~fs peaks around 5.3~eV, again in good agreement with the
measurement shown in Fig.~\ref{fig_proton_40fs10fs}.

%-----------------------------------
% Figure 12 : proton spectra = f(GDD)
\begin{figure}
\begin{center}
\includegraphics*[width=0.5\textwidth]{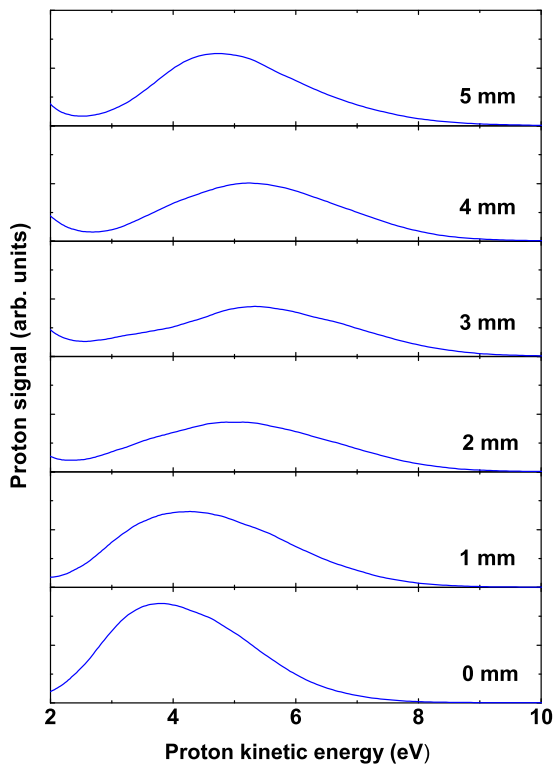}
\end{center}
\caption{(Color online) Optimization of the ultrashort pulse
duration. Proton spectra from H$_2$ recorded at
$2 \times 10^{15}$~Wcm$^{-2}$ with different thicknesses of fused
silica from 0 to 5~mm inserted in the beam path before the ion
spectrometer.}
\label{fig_proton_GDD}
\end{figure}
%----------
The above experimental and theoretical results give an	
\textit{a posteriori} justification of our experimental method
of the pulse optimization. Indeed, the optimization of the ultrashort
pulse duration can be performed \textit{in situ} inside the ion
spectrometer using the proton spectrum since it shifts to higher
energies as the pulse duration is decreased. After the initial pulse
duration optimization using the interferometric autocorrelator,
the proton spectra are recorded with different thicknesses
of fused silica inserted in the beam just before the ion
spectrometer entrance window. An optimum of the proton
spectrum to higher energies is systematically looked for. 
If such an optimum is not found, then additional negative
group delay dispersion is introduced in the compressor by adding
additional bounces onto the chirped mirrors.
Figure~\ref{fig_proton_GDD} represents the results of the
optimization procedure using fused silica plates with
thicknesses up to 5~mm. The maximum shift of the proton
spectra is found with a thickness of 3~mm and corresponds to
the shortest pulse duration available within our set--up
inside the ion spectrometer. Moreover, one can infer from 
Fig.~\ref{fig_proton_GDD} that the total number of detected protons
presents a minimum at the shortest pulse duration for a fused silica
thickness of 3~mm. Indeed the shortest pulse duration is
associated to a reduced internuclear range and hence to a
higher energy gap in order to reach the double ionization
H$^+$~+~H$^+$ threshold. In addition, charge resonance enhanced
ionization plays a minor role in comparison with larger
internuclear distances produced with longer pulse
durations. Finally, such a procedure was used with
other molecules such as N$_2$ and O$_2$ and represents a
straightforward method for pulse duration optimization
\cite{Baldit_2005}.

%----------------------------------------------
\subsection{Carrier--envelope phase dependence}
%
%-----------------
% Figure 13 = f(CEP)
\begin{figure}
\begin{center}
\includegraphics*[width=0.5\textwidth]{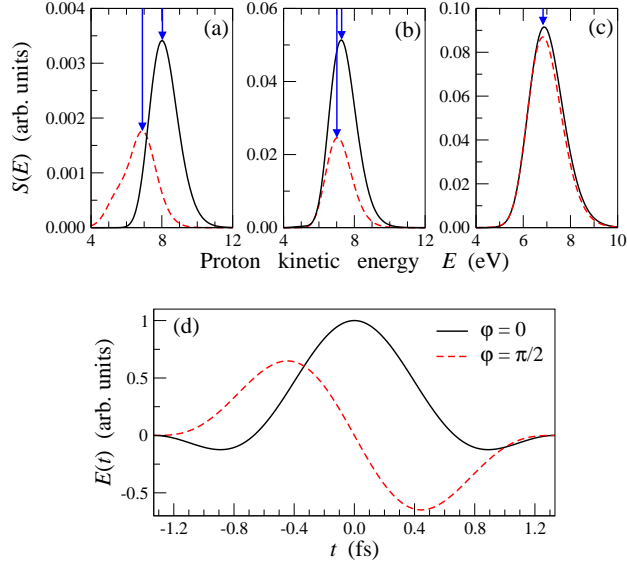}
\end{center}
\caption{(Color online) Calculated proton spectra from H$_2$
at $8 \times 10^{14}$~Wcm$^{-2}$ as a function of the
carrier--envelope phase $\varphi$ and the pulse duration.
Full curves: $\varphi = 0$. Dashed curves: $\varphi = \pi/2$.
The left (a), middle (b), and right (c) top panels correspond to
different pulse durations of respectively 1~fs, 2~fs, and 4~fs.
The vertical arrows mark the positions of the different peaks.
The corresponding electric field is represented in the lower
panel (d) as a function of time for a pulse duration of 1~fs.}
\label{fig_proton_CEP}
\end{figure}
%----------
%
The carrier--envelope phase dependence of strong field effects
was first investigated in high--order harmonic generation and
above--threshold ionization \cite{Baltuska_2003,Paulus_2003}.
Concerning molecular dissociation, Roudnev~\textit{et al.} have
theoretically shown that it is possible to control the H$_2^+$
and HD$^+$ dissociation with the carrier--envelope phase
\cite{Roudnev_2004}. More recently a first experimental evidence
was given in the dissociative ionization of D$_2$
\cite{Kling_2006}. The carrier--envelope phase dependence
of the intramolecular electronic motion leads to the localization
of the electron on one or the other nucleus. Therefore, the
resulting dissociation channels D$^+$~+~D and D~+~D$^+$, where
the left or right position of the deuteron D$^+$ indicates its
initial emission direction, can now be separated as a function of
the carrier--envelope phase. Finally, Tong and Lin investigated
the carrier-envelope phase dependence of nonsequential double
ionization of H$_2$ by few--cycle laser pulses \cite{Tong_2007}.
They found that the strong dependence of the double ionization
yields is due to the return energy of the rescattering electron.

Here we address the question of the carrier-envelope dependence
of the proton spectra from double ionization at laser intensities
where double ionization is mainly sequential. In the
nonsequential regime, Tong and Lin have shown that the proton
spectra lie within the same proton energy range \cite{Tong_2007}.
Moreover at laser intensities above
$1.5 \times 10^{14}$~Wcm$^{-2}$ and a pulse duration of 5~fs,
proton spectra from nonsequential double ionization are found to
be independent on the initial phase. We theoretically confirm
this tendency at higher laser intensities in the sequential
regime. Figure~\ref{fig_proton_CEP} represents proton spectra
calculated at $8 \times 10^{14}$~Wcm$^{-2}$ for two values of
the carrier-envelope phase $\varphi = 0$ and $\varphi = \pi/2$~rad.
In addition, calculations were performed for three values of
the laser pulse duration (a)~1~fs, (b)~2~fs, and (c)~4~fs. A
significant shift of 1.2~eV is only observed for the 1--fs pulse
whereas no energy shift is visible for a pulse duration of 4~fs.
In addition a smaller production of H$^+$ ions is calculated with
$\varphi = \pi/2$ as compared with $\varphi = 0$ for 1--fs and 2--fs
pulses. One can also notice in Fig.~\ref{fig_proton_CEP}(d) which
represents the time dependence of the associated electric fields
in the case of the 1--fs pulse, that a higher peak intensity is
achieved with $\varphi = 0$. In consequence, the sequential
ionization of the \htwop ion is more delayed at $\varphi = \pi/2$
than at $\varphi = 0$, thus resulting in a higher proton energy
when $\varphi = 0$. However as the number of cycles
increases within the pulse duration, this effect disappears
and the proton energy range is no more dependent on the
carrier--envelope phase.

%--------------------------------------
\subsection{Laser intensity dependence}
%
%----------------------------------------------
% Figure 14 : proton spectra = f(I) exp. and th.
\begin{figure}
\begin{center}
\includegraphics*[width=0.5\textwidth]{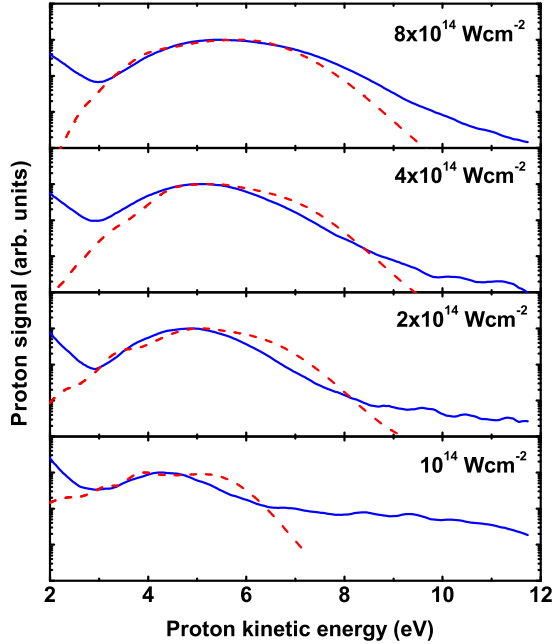}
\end{center}
\caption{(Color online) Proton spectra from H$_2$ at different laser
intensities $8 \times 10^{14}$, $4 \times 10^{14}$,
$2 \times 10^{14}$, and $10^{14}$~Wcm$^{-2}$ from top to bottom.
Full curves: Experimental. Dashed curves: Theoretical.}
\label{fig_proton_ecl_exp_th}
\end{figure}
%-----------
%
Double ionization of hydrogen by infrared laser light is a highly
nonlinear process. The intensity dependence therefore follows a
complicated behaviour which has to be studied both from
combined experimental and theoretical works.
Figure~\ref{fig_proton_ecl_exp_th} presents four experimental and
theoretical proton spectra from $10^{14}$ to
$8 \times 10^{14}$~Wcm$^{-2}$. The spectra are shifted towards
higher energies as the intensity is increased. The
experimental spectra are maximum at 4.2~eV, 4.9~eV, 5.0~eV,
and 5.5~eV at respectively $10^{14}$, $2 \times 10^{14}$,
$4 \times 10^{14}$, and $8 \times 10^{14}$~Wcm$^{-2}$.
This effect was also observed in multiple ionization of nitrogen
and oxygen with pulse durations of 40~fs and 10~fs and comes
from sequential electron emission
\cite{Quaglia_2002,Baldit_2005}. The time delay between the
first ionization and the second ionization, leading respectively
to H$_2^+$ and to H$^+$~+~H$^+$, is reduced because the necessary
instantaneous intensity for the second ionization comes sooner
when the laser intensity is increased. 
	
The experimental spectra are reasonably well reproduced by the model
predictions with some differences mainly in the high energy sides
of the spectra. In spite of these
discrepancies, the model predicts the correct energy range of the
expected proton spectra at 10~fs for different laser intensities.
It therefore constitutes a worthy predictive tool for the use
of ultrashort pulses when one considers the very complicated
dynamics of the hydrogen molecular response. Several effects
might explain the observed differences in addition to the laser
and ion measurements uncertainties. The calculation is performed
for one laser intensity and thus does not take into account the
intensity distribution within the focal volume. This might explain
the discrepancies in the high--energy wings of the spectra where
nonsequential double ionization does play a noticeable role at
low laser intensity \cite{Beylerian_2006}. Indeed for a given
peak laser intensity, low intensities are distributed over a
much larger volume than high intensities and may noticeably contribute
to the observed discrepancies in Fig.~\ref{fig_proton_ecl_exp_th}.

% CONCLUSION
\section{Summary and concluding remarks}
We have presented a detailed experimental and theoretical analysis
of \htwo sequential double ionization using ultrafast laser pulses
in the $10^{14}$--$10^{15}$~Wcm$^{-2}$ intensity range. We have shown
that the resulting proton spectra are very sensitive to the
temporal, phase, and intensity characteristics of the pulse. More
precisely, higher energy protons are emitted when the pulse
duration decreases and when the peak intensity increases.
On the other hand, the effect of the carrier--envelope phase
offset is only significant for pulse durations shorter than 4~fs.  

The main characteristics of the measured proton
spectra from double ionization of H$_2$ are well reproduced
by a full quantum calculation based on the time--dependent
Schr\"odinger equation. The $R$--dependent soft--Coulomb parameters
introduced in this study make the model relatively simple and
manageable using tractable computing facilities. These parameters are
fitted using \textit{ab initio} calculations of the ground electronic
states of H$_2$ and H$_2^+$ as a function of the internuclear
distance $R$. We are confident that the good agreement between
experiment and theory at 10~fs allows to give valuable
predictions for shorter laser pulses and in particular about the
ultrafast dynamics of the H$_2^+$ ion as a function of the pulse
duration, carrier--envelope phase offset and peak intensity.

The degree of control of few--cycle laser pulses becomes more and more
sophisticated \cite{Kling_2006}. In this respect, the H$_2$ response to the
associated high laser field might become a complementary diagnostic to
ultrashort laser pulses in addition to the existing FROG and SPIDER
techniques. In particular the relative simplicity of the proton
detection associated to a reliable model might be of some interest
in ultrafast laser physics.

%----------------------
\begin{acknowledgments}
We acknowledge high performance computing facilities of the IDRIS-CNRS
center (Project No. 06-1459) and financial support from ACI Photonique
Physique Attoseconde, from CEA via LRC-DSM Grant No. 05-33, and from
ANR Image Femto (Project No. BLAN07-2$\_\,$201076).
Laboratoire de Photophysique Mol\'eculaire is associated to
Universit\'e Paris-Sud.
\end{acknowledgments}
%--------------------
%
%===========
% REFERENCES
% PREPARATION : ``h2long07.bib'' used as a distinct ``*.bib'' file
% \bibliography{h2long07}
%
% SUBMISSION  : copy here ``h2ps05.bbl''

%
\end{document}